\documentclass[twoside,12pt]{article}

\usepackage{setspace}

\usepackage{fancyhdr}
\usepackage{amsfonts}
\usepackage{amsmath}
\usepackage{graphicx}
\usepackage{latexsym}

\newcommand{\fii}{\varphi}

\newcommand{\tsum}{\sum}

\begin{document}

\vbox{

\title{\large Translation of \\``Die Messung quantenmechanischer Operatoren"\\ by E.P.~Wigner}

\author{\normalsize P. Busch\\
\normalsize Department of Mathematics, University of York, York, YO10 5DD, UK\\
\normalsize Electronic address: paul.busch@york.ac.uk}

\date{}

\maketitle

\abstract{This is a {\em facsimile-style} translation of Wigner's seminal paper on measurement limitations
in the presence of additive conservation laws. A critical survey of the history of subsequent extensions and
variations of what is now known as the Wigner-Araki-Yanase (WAY) Theorem is provided in a paper published
concurrently.}

\thispagestyle{empty}
}

\newpage

\noindent Zeitschrift f\"ur Physik, Bd.~133, S.~101-108 (1952).

\vspace{12pt}

\begin{center}
{\bf \large Measurement of quantum mechanical operators.}

\vspace{6pt}
By\\
\vspace{6pt}
 {\large E.~P.~{\sc Wigner}.}\\
 \vspace{6pt}
 {\em (Received on 24 May 1952.)}
\end{center}


\renewcommand{\headrulewidth}{0pt}

\fancyhead[RO,LE]{\thepage}
\fancyhf[REH,LOH]{}        
\fancyhead[CE]{E.~P.~{\sc Wigner}:}
\fancyhead[CO]{Measurement of quantum mechanical operators.}
\fancyfoot[RO,LE]{}
\fancyfoot[RE,LO,C]{}

\setcounter{page}{101}

\thispagestyle{empty}

\begin{spacing}{0.8}
\noindent
{\small The usual assumption of the statistical interpretation of quantum mechanics that all hermitian operators represent
measurable quantities appears to be commonly recognized as a convenient mathematical idealization rather than a 
matter of fact. Here it is shown that already the validity of conservation laws for quantum magnitudes (such as the law of angular
momentum conservation or the law of electrical charge conservation), which govern the interaction between measured object and
measurement apparatus, allow the measurement of most operators only as a limit case. In particular, it is likely that the conditions for 
the measurement of operators that do not commute with the total charge cannot be satisfied. The same appears to be the case for
operators which do not commute with the number of baryons.}
\end{spacing}

\vspace{12pt}

1. The basic idea of the statistical interpretation of quantum mechanics was first stated by {\sc Born}\footnote{{\sc Born}, M.: Z.~Physik {\bf 37},
803 (1926). ``The motion of particles follows probability laws, the probability itself propagates in accordance with the law of causality."}
The intuitive-physical content of his ideas were elucidated and further elaborated in investigations by {\sc Heisenberg} and 
{\sc Bohr}\footnote{{\sc Heisenberg}. W.: Z.~Physik {\bf 43}, 172 (1927). --- Die Physikalischen Prinzipien der Quantenmechanik. Leipzig 1930. 
--- {\sc Bohr}, N.: Nature, Lond.~{\bf 121}, 580 (1928). --- Naturwiss.~{\bf 17}, 483 (1929) and further articles in the Max Planck--issue of
Naturwissenschaften. Cf.\ also {\sc Mott}, N.~F.:  Proc.~Roy.~Soc.~Lond.~{\bf 126}, 79 (1929) and {\sc Bohr}, N., and 
L.~{\sc Rosenfeld}: Phys.~Rev.~{\bf 78}, 794 (1950).} and others. The mathematical formalization of the theory is particularly due to the
work of von {\sc Neumann}\footnote{{\sc Neumann}, J.~v.: Mathematische Grundlagen der Quantenmechanik, esp.\ Chap.~VI. Berlin 1932.}.
It is a basic assumption of the theory that to every selfadjoint operator $Q$ corresponds a measurable physical quantity. The measurement
result is always an eigenvalue of the operator $Q$; at the same time, the measurement transforms the system into the state that is described
by the eigenfunction associated with the measurement result. Thus, let $\varphi$ be the original state function of the system and let us denote
the eigenvalues and eigenfunctions of $Q$ as  $q_1,q_2,\ldots$ and $\psi_1,\psi_2,\ldots$, respectively. Then the measurement yields, with
probability $|(\psi_\nu,\varphi)|^2$, the result $q_\nu$ and the system is in state $\psi_\nu$ after the measurement\footnote{If the eigenvalue
$q_\nu$ is degenerate and comprises several eigenstates $\psi_{\nu1},\psi_{\nu2},\ldots$, then the probability of $q_\nu$ is equal to 
$w_\nu=\tsum_\kappa|(\psi_{\nu\kappa},\varphi)|^2$ and the state after the measurement is 
$\tsum_\kappa w_{\nu}^{-1/2}(\psi_{\nu\kappa},\varphi)\psi_{\nu\kappa}$.}. Here $(\psi_\nu,\varphi)$ denotes the hermitian scalar product of
$\psi_\nu$ and $\varphi$. 
It may be noted at this point that an operator $I$ that leaves invariant  the transition probabilities as well as the temporal
change of the system also cannot influence the state itself. More specifically, if for all $\fii$ and $\psi$ both $|(\psi,\fii)|^2=|(\psi,I\fii)|^2$ and 
$(I\fii)_t=I(\fii_t)$ (where the index $t$ describes the temporal change of the system) then the states $\fii$ and $I\fii$ are altogether indistinguishable.
In the orthodox formulation of the theory the $I$ are complex numbers of absolute value 1.

The great weakness of the formalism sketched above is that it contains no prescription as to how the measurement of the operator $Q$ can be carried out.
It is easy to give such a prescription\footnote[1]{{\sc Neumann}, J.~v.: Mathematische Grundlagen der Quantenmechanik, esp.\ Chap.~VI.~Berlin 1932.}:
Combine the measured object $\fii$ with a measurement instrument, whose state function may be denoted $\xi$. The measuring instrument is
designed in such a way that the state function of the total system $\fii\xi$, consisting of object and instrument, changes into
\begin{equation}
\fii\xi\to\tsum_{\nu}(\psi_\nu,\fii)\psi_\nu\chi_\nu
\end{equation}
after a certain time, where the $\chi_\nu$ are the macroscopically distinguishable states of the measuring instrument: the state $\chi_\nu$ 
indicates the measurement result $q_\nu$~\footnote[2]{From (1) one can recognize the origin of the hermitian character of the operators corresponding
to observable quantities. The transition from the left hand side of (1) to the right is effected through a unitary operator. This operator sends $\psi_\nu\xi$ and
$\psi_\mu\xi$ to $\psi_\nu\chi_\nu$ and $\psi_\mu\chi_\mu$, respectively. But these latter functions are mutually orthogonal since the $\chi$, being
macroscopically distinguishable, must be mutually orthogonal. Due to the unitarity of the transition this follows then also for the $\psi_\nu\xi$, that is,
also for the $\psi_\nu$, which form the system of eigenfunctions of $Q$. Since also the $q_\nu$, being measurement results, are real numbers, the 
selfadjoint character of $Q$ follows.}. However, this prescription remains purely formal as long as it is not specified how the measurement instrument
is to be constructed in state $\xi$. A further, epistemologically even deeper, difficulty of the theory is glossed over with the words 
``macroscopically distinguishable states". This point was already extensively, and as exhaustively as presently possible, discussed by {\sc Heisenberg}
and will not be taken up here again. The question to be discussed concerns the possibility of an interaction between measurement object and measuring
instrument as symbolized by (1). Hence we will only obtain necessary conditions for the measurability of a quantity.  Even when an interaction
corresponding to (1) does not contradict any principle identified here, it is still well possible that either $\xi$ is not realizable as a matter of principle or
that the $\chi_\nu$ are not accessible to a direct or indirect macroscopic

\newpage
\noindent distinction. As a matter of fact\footnote[1]{Cf.~{\sc Heisenberg}, W.: l.c. and 
{\sc Neumann}, J.~v.: l.c., viz.~p.~223, 224.}, eq.~(1) only shifts the question of the distinguishability
of the $\psi_\nu$  to the distinguishability of the $\chi_\nu$, and here merely the conditions and possibility of such a shift will be investigated.

2. As long as one remains within the framework of the general theory, expressed by (1), no concrete statements can be made about the measurability
that would go beyond those of footnote 2 of the previous page. In fact, the measuring instrument of (1) could generally be a very simple and elementary system,
in an example due to {\sc Heisenberg} it consists of a single light quantum\footnote[2]{{\sc Heisenberg}, W.: l.c., Chapter II, 2. example b.}.
If, however, the postulate of relativistic invariance is brought into consideration, it should be widely known that there is at least one operator $I_1$ 
that is commutable with all observable quantities $Q$. The operator $I_1$ leaves all states with integer angular momentum unchanged but multiplies
 by $-1$ all states with half-integer angular momentum. The observation of a quantity whose operator does not commute with $I_1$ [such as, for instance,
 the quantized amplitudes $\psi(x,y,z)+\psi(x,y,z)^*$] would allow it to distinguish between states that must remain indistinguishable according to the
 theory of relativity\footnote[3]{This point will be discussed further in a somewhat popular way in a forthcoming article by {\sc Wick, Wightman} and {\sc Wigner}.
The present paper owes its origin to a problem that arose in the course of writing that article.}. This restriction of observability is independent of the theory
of measurement as expressed in (1). Here we want to address yet another kind of restriction that has its origin in the conservation laws of quantized magnitudes
and arises in a discussion of the possibility of the mapping  (1). This limitation will not be as strict as the one mentioned above and will merely entail that
the measuring instrument must be very large, in the sense that it must contain, with considerable probability, a very large amount of any quantized conserved
magnitude whose operator does not commute with its operator. 

3. Quantized conserved magnitudes of the above kind are, for example, the component of angular momentum in a specific direction, the total electric charge of the 
system, the number of ``heavy particles" therein. Henceforth the lower index of a state function will denote the number of quanta ($\hbar$, $e$, etc.) contained
in the state described by the state function. In order to avoid fractional indices, the index will be increased by $\frac 12$ where necessary. Furthermore (1)
will be adopted in its original form to begin with, other definitions of measurement will be discussed at the end of this paper.

\newpage
\vbox{
In the simplest case the eigenfunctions of a typical operator that does not commute with the conserved quantity have the form 
$(\psi_0+\psi_1)/\sqrt 2$ and $(\psi_0-\psi_1)/\sqrt 2$. If, for example, the conserved quantity is the angular momentum in the $Z$-direction,
$\psi_0+\psi_1$ and $\psi_0-\psi_1$ are eigenfunctions of the $X$-components of the spin of a particle. The operator associated with this component 
obviously does not commute with  the angular momentum in the $Z$-direction. Eq.~(1) thus reads
\begin{equation}
\left.\begin{split}
(\psi_0+\psi_1)\xi\to(\psi_0+\psi_1)\chi\\
(\psi_0-\psi_1)\xi\to(\psi_0-\psi_1)\chi',
\end{split}\right\}
\end{equation}
where $(\chi,\chi')=0$ and the arrow represents a linear unitary transformation that commutes with the operator of the conserved quantity.
If we add and subtract the two eqs.~(2) and simultaneously decompose $\chi+\chi'$ and $\chi-\chi'$ into the parts $\sigma_\nu$ and $\tau_\nu$, 
respectively, which represent states with a sharply determined value of the conserved quantity,
\begin{equation}
(\chi+\chi')/\sqrt 2=\tsum\sigma_\nu,\qquad (\chi-\chi')/\sqrt 2=\tsum\tau_\nu,
\end{equation}
we obtain
\vspace{-10pt}
\begin{subequations}
\begin{align}
\psi_0\xi\to(\psi_0\tsum\sigma_\nu+\psi_1\tsum\tau_\nu)/\sqrt 2,\\
\psi_1\xi\to(\psi_0\tsum\tau_\nu+\psi_1\tsum\sigma_\nu)/\sqrt 2.
\end{align}
\end{subequations}
\vspace{-18pt}

\noindent
It is already evident from (4) that the measuring apparatus must contain an infinite amount of the conserved quantity. According to (4),
the expectation value of the conserved quantity is the same for the two states represented by the right hand sides of (4). Its expectation
value for the measured object is equal to $\frac 12$ in both cases, its expectation value for the measuring instrument is in both cases 
the arithmetic mean of the expectation values for $\tsum\sigma_\nu$ and $\tsum\tau_\nu$. Indeed, after the measurement the measured
object and measuring instrument are again separated and the total content of the conserved quantity in the system is composed additively
of the contents of the measured object and measuring instrument. By contrast, the expectation value of the conserved quantity is greater
by 1 for the left hand side of (4b) than (4a).

One can sharpen the contradiction if one notes that from (4) and the conservation law  the equations
\vspace{-10pt}
\begin{equation}
\left.\begin{split}
\psi_0\xi_\nu&\to(\psi_0\sigma_\nu+\psi_1\tau_{\nu-1})/\sqrt2\\
\psi_1\xi_{\nu-1}&\to(\psi_0\tau_\nu+\psi_1\sigma_{\nu-1})/\sqrt2
\end{split}\right\}
\end{equation}
\vspace{-10pt}

\noindent
follow, where the $\xi_\nu$ are the components of $\xi$ in terms of eigenfunctions of the conserved quantity,
\begin{equation}
\xi=\tsum\xi_\nu.
\end{equation}
If we now introduce the notation
\vspace{-10pt}
\begin{equation}
(\xi_\nu,\xi_\nu)=x_\nu;\quad (\sigma_\nu,\sigma_\nu)=s_\nu;\quad (\tau_\nu,\tau_\nu)=t_\nu;\quad 
(\sigma_\nu,\tau_\nu)=a_\nu+ib_\nu
\end{equation}
}
\noindent ($x_\nu,s_\nu,t_\nu,a_\nu,b_\nu$ real), the equations
\vspace{-6pt}
\begin{subequations}
\begin{align}
x_\nu={\tfrac 12}s_\nu&+{\tfrac 12} t_{\nu-1},\qquad x_{\nu-1}={\tfrac 12}t_\nu+{\tfrac 12}s_{\nu-1},\\
&0=a_\nu-ib_\nu+a_{\nu-1}+ib_{\nu-1}
\end{align}
\end{subequations}
\vspace{-18pt}

\noindent
express the unitary character of the transition indicated by the arrow in (5),
\vspace{-10pt}
\begin{equation}
\tsum x_\nu=\tsum s_\nu=\tsum t_\nu=1,\qquad \tsum a_\nu=\tsum b_\nu=0
\end{equation}
the normalization of $\xi$, $\chi$, $\chi'$ and the orthogonality of $\chi$ and $\chi'$. But  (8b) and (9)
immediately entail that \hbox{$a_\nu=b_\nu=0$}, from (8b) it follows that
\vspace{-8pt}
\begin{equation*}
x_{\nu+1}-\tfrac 12 s_{\nu+1}=\tfrac 12 t_\nu=x_{\nu-1}-\tfrac 12s_{\nu-1},
\end{equation*}
\vspace{-20pt}

\noindent
so that $x_{2\nu+1}-\frac 12s_{2\nu+1}$ as well as $x_{2\nu}-\frac 12 s_{2\nu}$ are independent of $\nu$.
This holds then also for $t_\nu$, which, however, is incompatible with (9). Hence, strictly speaking, a measurement
that leads to a separation of $\psi_0+\psi_1$ and $\psi_0-\psi_1$ is impossible. With the help of somewhat more
tedious algebra, which however is not significantly different from the above,  the same can also be shown for the
states  $\alpha\psi_0+\beta\psi_1$ and $-\bar\beta\psi_0+\bar\alpha\psi_1$, where $\alpha$ and $\beta$ are arbitrary
complex numbers.

Since a measurement of the spin components is practically possible, it must also be possible to modify the preceding
consideration in such a way that it demonstrates the possibility  of such a measurement with arbitrary accuracy. To this end,
let us denote the states into which $(\psi_0+\psi_1)\xi$  and $(\psi_0-\psi_1)\xi$  are transformed through the 
measurement process as
\vspace{-6pt}
\begin{equation}
\left.\begin{split}
(\psi_0+\psi_1)\xi&\to(\psi_0+\psi_1)\chi+(\psi_0-\psi_1)\eta\\
(\psi_0-\psi_1)\xi&\to(\psi_0-\psi_1)\chi'+(\psi_0+\psi_1)\eta'.
\end{split}\right\}
\end{equation}
\vspace{-12pt}

\noindent
Now if $(\chi,\chi')=0$ remains and $(\eta,\eta)$ and $(\eta',\eta')$  can be made arbitrarily small, then by determination
of the state $\chi$ resp. $\chi'$ of the measuring instrument one can infer the state of the measured object in almost all cases.

We will arrange that  $\eta=-\eta'$ and also $(\eta,\chi)=(\eta,\chi')=(\chi,\chi')=0$. This means that the measurement can have
three results: the state is $(\psi_0+\psi_1)/\sqrt2$, the state is $(\psi_0-\psi_1)/\sqrt2$, the state is undetermined. But the 
probability for obtaining the last result is $(\eta,\eta)$, and as we will see shorly, this can be made arbitrarily small. To achieve this,
the expansion of $\xi$ according to (6) must, however,  have very many components.

We assume that this number is $n$ and that the measuring instrument can contain no less than one and no more than $n$ units of the
conserved quantity. 
 Then the $\xi_\nu$ vanish except for $0< \nu\le n$. Furthermore we introduce the abbreviations
 
 \newpage
 
 
 \vbox{
\begin{equation}
\left.\begin{split}
2\chi&=2\sigma+\rho+\tau\\
2\chi'&=2\sigma-\rho-\tau\\
2\eta&=-2\eta'=\tau-\rho
\end{split}\right\}
\end{equation}
 \vspace{-12pt}
 
 \noindent
This gives from (10)
\begin{equation}
\left.\begin{split}
\psi_0\xi&\to\psi_0\sigma+\psi_1\rho\\
\psi_1\xi&\to\psi_0\tau+\psi_1\sigma.
\end{split}\right\}
\end{equation}
The $\sigma,\tau,\rho$ can then, similarly to (6), be written as a sum of eigenfunctions of the conserved quantity.
Among the $\sigma_\nu$ only those with $0<\nu\le n$ are finite, but $\rho_0$ remains finite whereas $\rho_n$ vanishes already.
Conversely $\tau_1$ vanishes while $\tau_{n+1}$ is finite. 

The orthogonality of the right hand sides of (12) leads to
 \vspace{-6pt}
\begin{equation}
(\sigma_\nu,\tau_\nu)+(\rho_{\nu-1},\sigma_{\nu-1})=0,
\end{equation}
 \vspace{-18pt}
 
 \noindent
the normalization condition is
 \vspace{-6pt}
$$
(\xi_\nu,\xi_\nu)=(\sigma_\nu,\sigma_\nu)+(\rho_{\nu-1},\rho_{\nu-1})=(\sigma_\nu,\sigma_\nu)+(\tau_{\nu+1},\tau_{\nu+1}).\eqno{\rm (13a)}
$$
 \vspace{-16pt}
 
 \noindent
In addition there are the conditions
 \vspace{-6pt}
\begin{subequations}
\begin{align}
(\xi,\xi)&=\tsum (\xi_\nu,\xi_\nu)=1,\\
(\chi,\chi')&=4\tsum(\sigma_\nu,\sigma_\nu)-\tsum(\rho_\nu+\tau_\nu,\rho_\nu+\tau_\nu)=0,
\end{align}
\end{subequations}
 \vspace{-18pt}
 
 \noindent
and because of $(\chi,\eta)=(\chi',\eta)=0$
 \vspace{-6pt}
\begin{align*}
\tsum(\sigma_\nu,\tau_\nu-\rho_\nu)&=0\tag{14c}\\
\tsum(\tau_\nu+\rho_\nu,\tau_\nu-\rho_\nu)&=0.\tag{14d}
\end{align*}
 \vspace{-18pt}
 
 \noindent
One can satisfy these equations in manifold ways. The simplest choice --- which however does not lead to the smallest possible value
of $(\eta,\eta)$ --- is perhaps that for which all
 \vspace{-10pt}
\begin{equation}
(\sigma_\nu,\tau_\nu)=(\sigma_\nu,\rho_\nu)=0
\end{equation}
 \vspace{-18pt}
 
 \noindent
vanish. With this one has fulfilled (13) and (14c). 
Then, for those $\nu$ for which $\rho_\nu$ as well as $\tau_\nu$ can be finite one can assume
\vspace{-10pt}
\begin{align*}
\rho_\nu=\tau_\nu\quad (1<\nu\le n-1)\tag{15a}
\end{align*}
\vspace{-18pt}

\noindent
and give all nonvanishing $\rho,\tau$
\vspace{-10pt}
\begin{align*}
(\rho_\nu,\rho_\nu)=(\tau_\nu,\tau_\nu)=c'\tag{15b}
\end{align*}
\vspace{-18pt}

\noindent
the same norm. Then (14d) is satisfied and also (13a) if one uses it for the determination of the $(\xi_\nu,\xi_\nu)$.

Finally one can also assume
\begin{align*}
(\sigma_\nu,\sigma_\nu)=c\tag{15c}
\end{align*}
}
\noindent
independently of $\nu$ (for $0<\nu\le n$). It follows then also that $(\xi_\nu,\xi_\nu)=c+c'$ and because of (14a)
\vspace{-10pt}
\begin{align*}
n(c+c')=1.\tag{16a}
\end{align*}
\vspace{-18pt}

\noindent
It remains only to satisfy (14b). This gives
\begin{equation*}
\left.\begin{split}
4nc=&(\rho_0,\rho_0)+(\rho_1,\rho_1)+(\tau_n,\tau_n)+(\tau_{n+1},\tau_{n+1})+\\
&\quad+\tsum_{\nu=2}^{\nu=n-1}(2\rho_\nu,2\rho_\nu)=4c'+4(n-2)c')=4(n-1)c'.
\end{split}\right\}\tag{16b}
\end{equation*}
From (16a) and (16b) one computes $c'=1/(2n-1)$. If one finally computes $(\eta,\eta)$, the terms with 
$\nu=2,3,\ldots,n-1$ drop out because of (15a) and one obtains
\begin{equation*}
(\eta,\eta)=c'=1/(2n-1).\tag{17}
\end{equation*}
This approaches zero indeed if $n$ becomes very large. Through a more favorable choice of $\sigma,\tau,\rho$ one
could have achieved that $(\eta,\eta)$ approaches zero as $1/n^2$. Nevertheless $\xi$ will need to have a very large
number of components, hence the measuring apparatus a very large amount of the conserved quantity, if one wants to
have a high level of confidence that the interaction between the measured object and the measuring apparatus leads to
a measurement. In particular, if one wants to measure the phase difference between parts of the state function that correspond to
different total charges\footnote[1]{This point will be discussed further in a somewhat popular way in a forthcoming article by {\sc Wick, Wightman} and {\sc Wigner}.
The present paper owes its origin to a problem that arose in the course of writing that article.}, the electrical charge of the measuring
apparatus must be largely indeterminate --- if such a measurement is  at all possible.

4. The question remains whether the description of a measurement contained in (1) or (2) is too demanding. It will, however, become 
apparent that although this is probably the case, even a substantially relaxed definition of measurement leads to similar results.

The most important generalization of (2) probably consists in allowing a change of the state of the measured object  even when originally it
was in one of the two states $\psi_0+\psi_1£$ or $\psi_0-\psi_1$. If the measurement is required to merely distinguish between these two
states, the final state of the measured object will indeed remain irrelevant (cf.\ also several of the examples in footnote 2 at the beginning).
It was already noted above that even the determination of a difference between $\fii+\fii'$ and $\fii-\fii'$ is prohibited if $\fii$ describes a state
with integer and $\fii'$ a state with half-integer angular momentum.

If we alter (2) so as to simply replace $\psi_0$ and $\psi_1$ by $\psi_0'$ and $\psi_1'$, respectively, on the right hand side, nothing changes in
the preceding considerations. 
Indeed the solvability of the equations thus obtained would also entail the solvability of (2) in their original form.
Therefore we want to assume 
\begin{equation*}
\left.\begin{split}
(\psi_0+\psi_1)\xi&\to(\tsum\psi_\mu')(\tsum \chi_\lambda')\\
(\psi_0-\psi_1)\xi&\to(\tsum\psi_\mu'')(\tsum \chi_\lambda'').
\end{split}\right\}\tag{18}
\end{equation*}
generally. However we restrict ourselves to the case where the number of quanta of the conserved quantity in $\xi$ is definite. This number
can be assumed to be equal to zero without loss of generality.

Since the left hand sides of (18) contain either no or one quantum of the conserved quantity it follows that
\begin{equation*}
\tsum_\mu \psi_\mu'\chi_{\nu-\mu}'=0,\qquad \nu\neq 0,1.\tag{19}
\end{equation*}
Due to the orthogonality of the terms of the sum in (19) they must all vanish individually. It can be assumed again $\psi_0'$ and $\chi_0'$ are finite
and consequently only the following two possible cases remain:\\
\phantom{nnnn}1. $\psi_0',\psi_1',\chi_0'$ finite, all other vanish;\\
\phantom{nnnn}2.  $\psi_0',\chi_0',\chi_1'$ finite, all other vanish.\\
For the double-primed quantities it holds as well that either only two $\psi''$ and one $\chi''$ or only one $\psi''$ and two $\chi''$ can be finite.
In addition it follows from case 1 that
\begin{equation*}
2\psi_0\xi\to(\psi_0'+\psi_1')\chi_0'+\tsum\psi_\mu''\chi_\lambda''
\end{equation*}
that $\psi_1''\chi_0''$ must be finite and, in fact, equal to $-\psi_1'\chi_0'$. Likewise in case 2 it follows that $\psi_0'\chi_1'=-\psi_0''\chi_1''$.
A very simple discussion now leads to the result that case (1) entails the modification of (2) that has already been considered in the preceding
section. By contrast, case 2 leads essentially to
\begin{equation}
\left.\begin{split}
(\psi_0+\psi_1)\xi\to\psi_0'(\chi_0+\chi_1)\\
(\psi_0-\psi_1)\xi\to\psi_0'(\chi_0-\chi_1)
\end{split}\right\}\tag{20}
\end{equation}
instead of (2). In this case the measuring process leads to an exchange of the conserved quantity between measured object and measuring
instrument. In particular the problem of the distinction between $\psi_0+\psi_1$ and $\psi_0-\psi_1$ is replaced by the almost equivalent
problem of the distinction between $\chi_0+\chi_1$ and $\chi_0-\chi_1$. Hence, if this difference is not directly apperceptible the result of the 
preceding section remains valid without modification.

\vspace{12pt}

\begin{center}
{\em Princeton (N.~J.)}, Palmer Physical Laboratory, University. 
\end{center}

\newpage

\thispagestyle{empty}



\section*{\normalsize Translator's Notes}

1. The book by Heisenberg cited in footnote 1 is available in English translation (by Carl Eckart and F.C.~Hoyt): ``The Physical Principles of the Quantum Theory",
University of Chicago Press, 1930, reprinted in Dover Publications, 1949.

\noindent
2. Footnote 3 of page 101: English translation: ``Mathematical Foundations of Quantum Mechanics", translated by R.~T.~Beyer, Princeton University Press, 1955.

\noindent
3. Footnote 1 of page 103: The corresponding page numbers in the English translation of von Neumann's book are pp.~419-421.

\noindent
4. Footnote 3 of page 103 refers to the paper by G.~C.~Wick, A.~S.~Wightman and E.~P.~Wigner, ``The Intrinsic Parity of Elementary Particles",
{\em Phys.~Rev.}~{\bf 88}, 101 (1952). Curiously, literally the same footnote appears at the bottom of page 107.

\noindent
5. Typographical error: the displayed unnumbered equation after Eq.~(9), p.~105, follows from (8a), not (8b) as stated in the text.


\vspace{12pt}

\noindent{\bf Acknowledgements.} Thanks are due to Leon Loveridge and Daniel McNulty for their careful reading of this translation and for valuable
linguistic comments.

\end{document}